\documentclass{PoS}

\usepackage{amsmath,wrapfig}
\usepackage[authoryear,square]{natbib}
\bibpunct{(}{)}{;}{a}{}{,}

\newcommand{\om}{\Omega_m}

\newcommand{\ho}{H_0}

\newcommand{\fnl}{{f_\mathrm{NL}}}
\newcommand{\fov}{\mathrm{FoV}}

\title{Cosmology on the Largest Scales with the SKA}

\ShortTitle{Cosmology on the Largest Scales}

\author{\speaker{Stefano Camera},$^{ab}$ Alvise Raccanelli,$^{cde}$ Philip Bull,$^f$ Daniele Bertacca,$^g$ Xuelei Chen,$^h$ Pedro G. Ferreira,$^i$ Martin Kunz,$^{jk}$ Roy Maartens,$^{gl}$ Yi Mao,$^m$ M\'ario G. Santos,$^{gnb}$ Paul R. Shapiro,$^o$ Matteo Viel$^{pq}$ and Yidong Xu$^h$\\
$^a$Jodrell Bank Centre for Astrophysics, University of Manchester, Manchester M13 9PL, UK; $^b$CENTRA, IST, Universidade de Lisboa, 1049-001 Lisboa, Portugal; $^c$Department of Physics \& Astronomy, Johns Hopkins University, Baltimore, MD 21218, USA; $^d$Jet Propulsion Laboratory, California Institute of Technology, Pasadena, CA 91109, USA; $^e$California Institute of Technology, Pasadena, CA 91125, USA; $^f$Institute of Theoretical Astrophysics, University of Oslo,  0315 Oslo, Norway; $^g$Department of Physics, University of the Western Cape, Cape Town 7535, South Africa; $^h$National Astronomical Observatories, Chinese Academy of Sciences, Beijing 100012, China; $^i$Astrophysics, Physics Department, University of Oxford, Oxford OX1 3RH, UK; $^j$D\'epartement de Physique Th\'eorique and Center for Astroparticle Physics, Universit\'e de Gen\`eve,  CH 1211 Gen\`eve 4, Switzerland; $^k$African Institute for Mathematical Sciences, Cape Town 7945, South Africa; $^l$Institute of Cosmology \& Gravitation, University of Portsmouth, Portsmouth P01 3FX, UK; $^m$Institut d'Astrophysique de Paris, Institut Lagrange de Paris, CNRS, UPMC Univ. Paris 06, UMR7095,  F-75014 Paris, France; $^n$SKA SA, 4th Floor, The Park, Park Road, Pinelands 7405, South Africa; $^o$Department of Astronomy and Texas Cosmology Center, University of Texas at Austin,  Austin, TX 78712, USA; $^p$INAF, Astronomical Observatory of Trieste, via Tiepolo 11, 34131 Trieste, Italy; $^q$INFN, Sezione di Trieste, 34100 Trieste, Italy\\
E-mail: \email{stefano.camera@manchester.ac.uk}}

\normalsize
\abstract{The study of the Universe on ultra-large scales is one of the major science cases for the Square Kilometre Array (SKA). The SKA will be able to probe a vast volume of the cosmos, thus representing a unique instrument, amongst next-generation cosmological experiments, for scrutinising the Universe's properties on the largest cosmic scales. Probing cosmic structures on extremely large scales will have many advantages. For instance, the growth of perturbations is well understood for those modes, since it falls fully within the linear r\'egime. Also, such scales are unaffected by the poorly understood feedback of baryonic physics. On ultra-large cosmic scales, two key effects become significant: primordial non-Gaussianity and relativistic corrections to cosmological observables. Moreover, if late-time acceleration is driven not by dark energy but by modifications to general relativity, then such modifications should become apparent near and above the horizon scale. As a result, the SKA is forecast to deliver transformational constraints on non-Gaussianity and to probe gravity on super-horizon scales for the first time.}

\FullConference{
Advancing Astrophysics with the Square Kilometre Array\\
June 8-13, 2014\\
Giardini Naxos, Italy}

\newcommand{\skipthis}[1]{}

\begin{document}

\section{Introduction}\label{sec:introduction}
An uncharted area of physical and observational cosmology is the physics of ultra-large scales. By this we mean length scales which are near or beyond the cosmic horizon. Most large-scale surveys have been limited to sampling length scales of order hundreds of megaparsecs, whilst deeper surveys (i.e. higher redshift surveys) are not yet good enough to accurately measure modes which are of order the horizon. Yet, the cosmological information hidden in the extremely large scales is of great importance for our comprehension of the Universe. For a start, the growth of cosmological perturbations is well within the linear regime on those ultra-large scales. This allows for better checks of our theoretical model, since we can there safely disregard the non-linear growth of structures---which always requires some degree of \textit{ad hoc} modelling. Furthermore, on very large scales the most relevant physics is simply given by the gravitational interaction, thus freeing us from the necessity of taking 
astrophysical processes into account.

Theoretical knowledge of the Universe on the largest cosmic scales can be very precise, but ultra-large scales have been most difficult to access. This is mainly due to two observational issues. Firstly, if we want to probe  very large scales---i.e.\ very large wavelengths or very small Fourier momenta---a wide field of view is not sufficient. For a wide-field, shallow survey the transverse components, $\mathbf k_\perp$, of the physical wavevector $\mathbf k$ may be arbitrarily small, but the smallest  measured wavenumber, $k\equiv|\mathbf k|=\sqrt{\mathbf k_\perp^{2}+k_\parallel^{2}}$, is limited by the fact that the survey probes only a thin shell---thus providing us with a large parallel component, $k_\parallel$. In other words, we need to observe a very large `cube' of Universe. This is a challenging task for conventional galaxy redshift surveys, for it is extremely hard to achieve the required sensitivity at high redshift over a large area of the sky. Closely related to this is the second observational 
problem, namely cosmic variance. Even though the full SKA will be able to detect HI galaxies as far away as $z\simeq2$ over roughly three quarters of the celestial dome---thus actually observing a huge cosmic cube---the relatively small number of sources with ultra-large separation can heavily limit the constraining potential of the survey.

The SKA is in a unique position to change this state of affairs and push forward the study of ultra-large scale modes. To tackle successfully the two observational issues described above, the SKA will employ two newly developed techniques, one being HI intensity mapping (IM) and the other the so-called multi-tracer approach. Since we can exploit IM both during and after the epoch of reionisation (EoR), the SKA will be able to probe exceptionally large redshifts. When mapping large-scale structure after the EoR, it is possible to efficiently collect information on large wavelengths without having to worry about resolving small-scale features or individual galaxies. When mapping the EoR, one can effectively sample a large number of horizon-size volumes at the time. 

In addition, the multi-tracer technique will allow us to beat down cosmic variance. This approach is based on galaxies being biased but not stochastic tracers of the underlying density field. Thus, by observing different galaxy populations---namely, different tracers---we can construct several realisations of the clustered halo distribution. Although measurements of global cosmological properties like the total matter density $\om$ or the Hubble constant $\ho$ will still be affected by cosmic variance, we will be able to constrain halo properties to a much higher degree of accuracy.

In this Chapter, we outline the science to be done by using the SKA to map the matter distribution on the largest cosmic scales. We first make the case for why ultra-large scales are interesting, focussing on relativistic effects, modified gravity (MG) and scale-dependent biasing arising from primordial non-Gaussianity (PNG). Then, we address how different methods will constrain these various physical phenomena, showing that indeed the SKA will be transformational for constraining PNG, and unique for probing gravity on super-horizon scales. Lastly, we discuss the issues that need to be faced when trying to probe the largest scales. We refer the reader to Chapters \citet{Santos:SKA14},  \citet{Abdalla:SKA14} and \citet{Jarvis:SKA14} for details on the phases of the SKA, its observation modes and the different methods we are discussing here.

\section{The Importance of Probing Ultra-Large Cosmic Scales}\label{sec:uls1}
The study of the biggest volumes ever of cosmic large-scale structure will allow a major advance in tackling two of the most fundamental questions in contemporary cosmology: What is the physical mechanism that generated the inflationary expansion in the early Universe? Does general relativity (GR) hold on the largest scales? 

Most models of inflation do predict some degree of non-Gaussianity in the distribution of primordial density fluctuations, but the simplest models predict negligible PNG. A detection of PNG will be vital for ruling out classes of models and advancing our understanding of inflation. Currently, the most stringent constraints on PNG come from the Planck satellite, but future cosmic microwave background (CMB) experiments are unlikely to improve these results significantly. The new frontier of constraining PNG is large-volume surveys of the matter distribution.

Similarly, tests of GR on cosmological scales are based on observations of the large-scale structure. Current constraints are weak, but with its huge surveyed volumes and multiple probes, the SKA will take the lead in the next generation of tests. Additionally, we can tighten the current constraints on dark energy (DE) and MG models by including much larger scales, thus increasing the statistical power of the observations and improving the sensitivity to any scale-dependent deviation from GR.

In this Section we review such ultra-large scale phenomena, highlighting their importance for our understanding of the Universe.

\subsection{General Relativistic Effects}\label{ssec:gr}
Probing ultra-large scales involves a theoretical challenge that has been recognised only recently. GR effects arise from observing on the past light-cone, which distorts the number counts and brightness temperature fluctuations on very large scales. So far, most analyses have been performed using a  Newtonian approximation, with the flat-sky redshift-space distortions (RSDs) grafted on as a special relativistic effect. This is the `standard' relativistic correction to the Newtonian approximation, which is significant on sub-horizon scales. Sometimes, the effect of lensing convergence is also included. This is adequate for past and present surveys, which analyse galaxy clustering on scales well below the horizon. Sometimes this Newtonian-like approach also includes the contribution of weak lensing magnification to the matter over-density. The lensing effects can be significant on small scales. But future wide and deep surveys will need to employ a more precise modelling, accounting not only for RSDs and lensing, but for all geometric and relativistic corrections. The full relativistic analysis includes terms that are suppressed on sub-horizon scales such as velocity (or Doppler) terms, Sachs-Wolfe (SW) and integrated SW (ISW) type terms, and time-delay contributions \cite[e.g.][]{Yoo:2010ni,Jeong:2011as,Bonvin:2011bg,Challinor:2011bk,Bruni:2011ta}.

One can work with the redshift-space correlation functions and include all GR and wide-angle contributions \citep{Bertacca:2012tp,Raccanelli:2013dza,Raccanelli:2013gja,Bonvin:2013ogt}. By analysing this, we can potentially extract more information about the structure of galaxy clustering. For instance, the odd Legendre multipoles of the correlation function vanish in the Newtonian flat-sky approximation, whereas they are in general non-zero for wide-angle separations and in the full GR analysis \citep{Raccanelli:2013dza,Bonvin:2013ogt}. Alternatively, the use of the angular power spectrum also avoids the flat-sky assumption, which is important for full-sky surveys with many wide-angle correlations \citep{Bonvin:2011bg,Challinor:2011bk,DiDio:2013bqa}.

\begin{figure}
\centering
\includegraphics[width=0.5\textwidth]{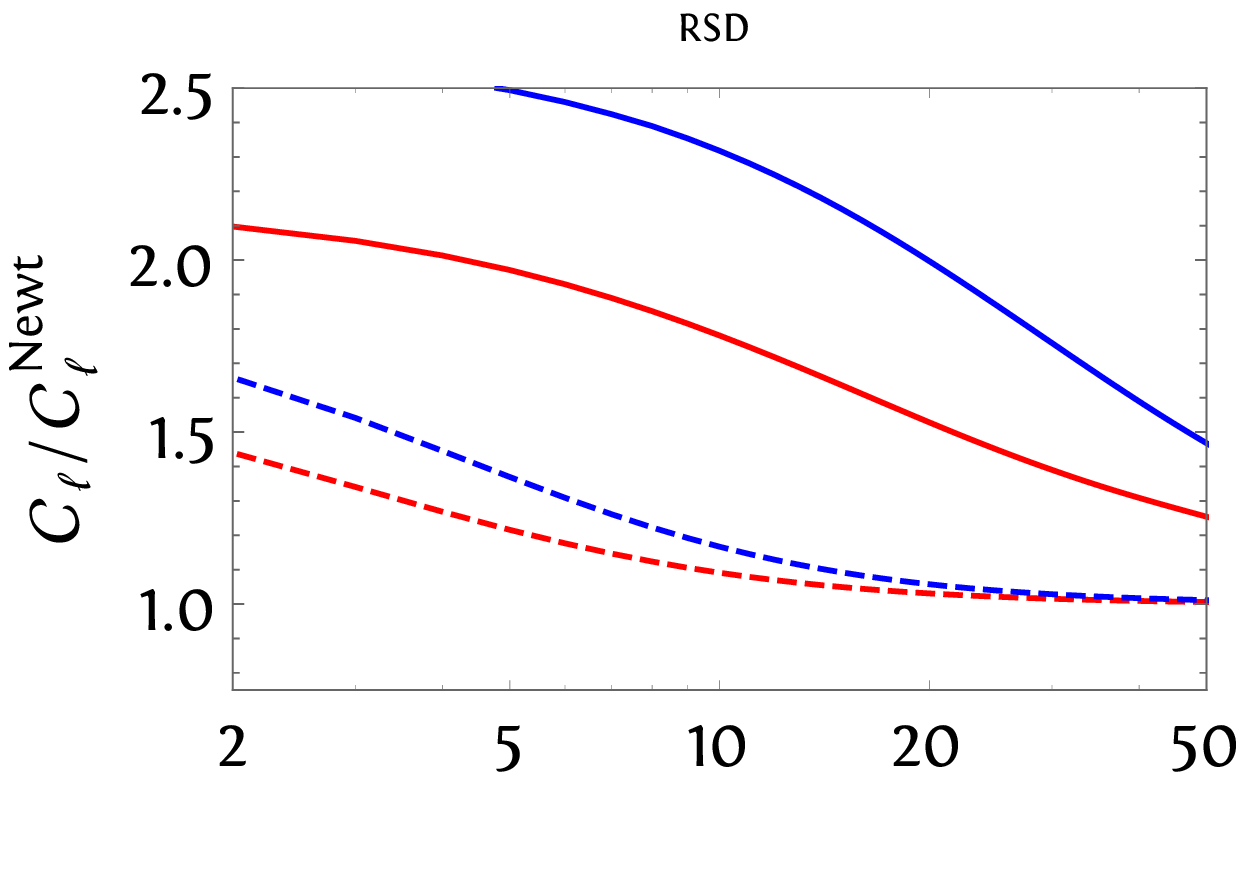}\includegraphics[width=0.5\textwidth]{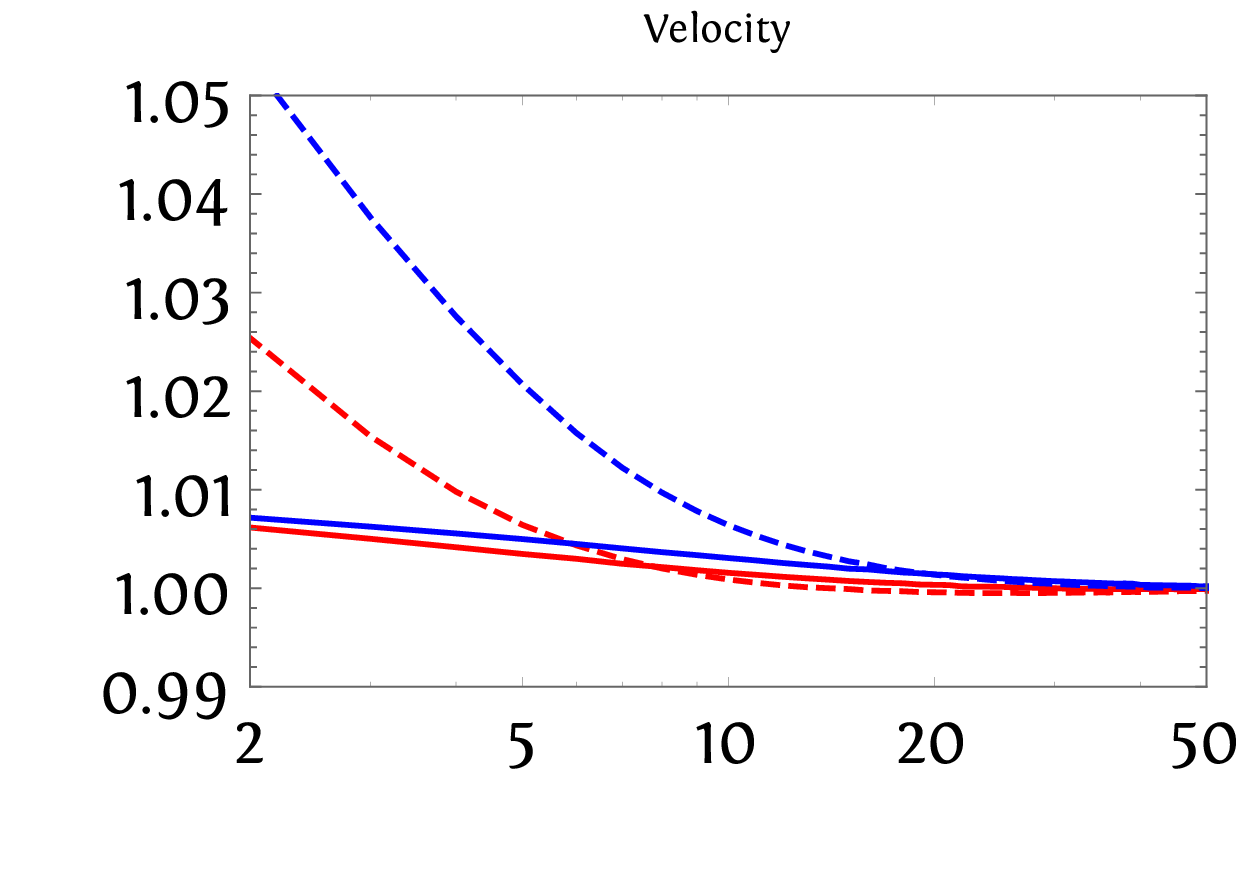}
\\\vskip -0.2in\includegraphics[width=0.5\textwidth]{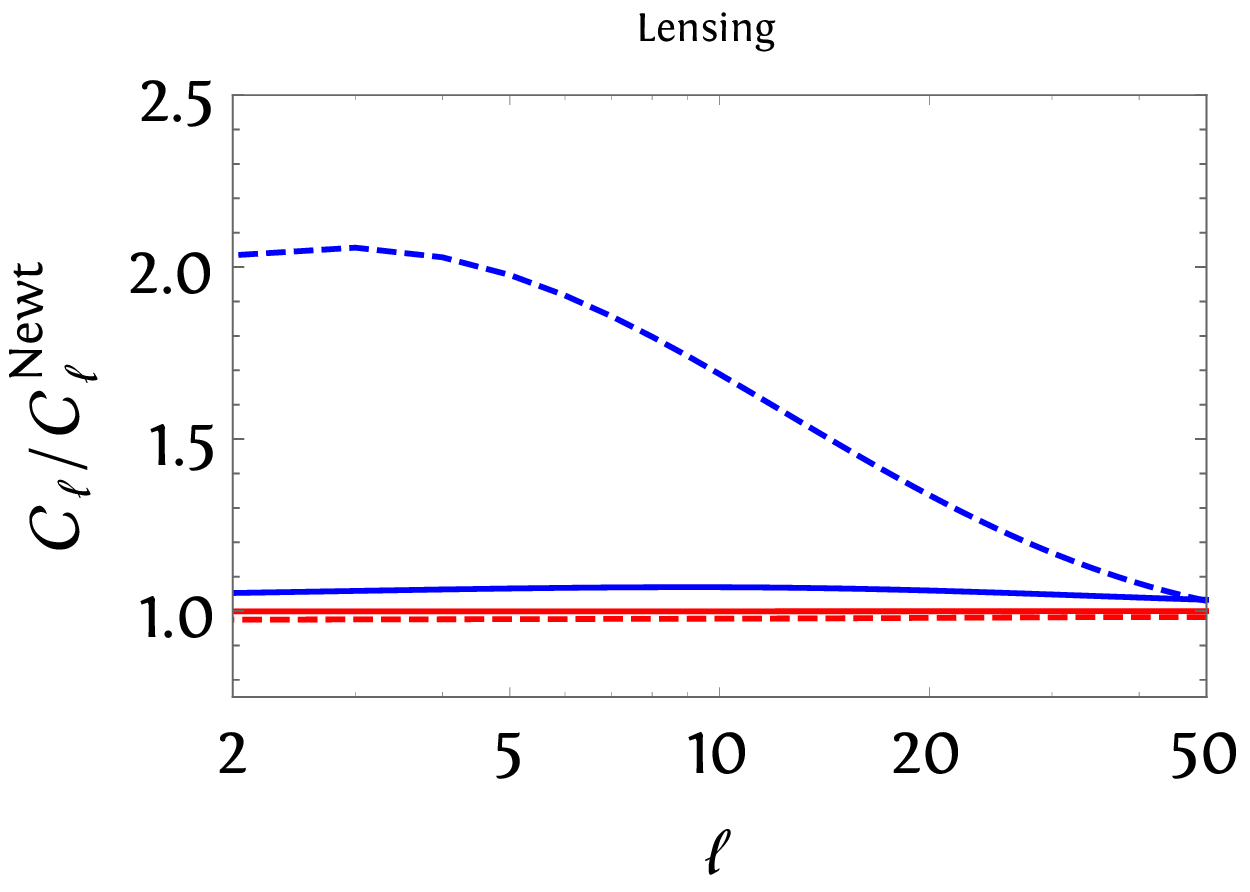}\includegraphics[width=0.5\textwidth]{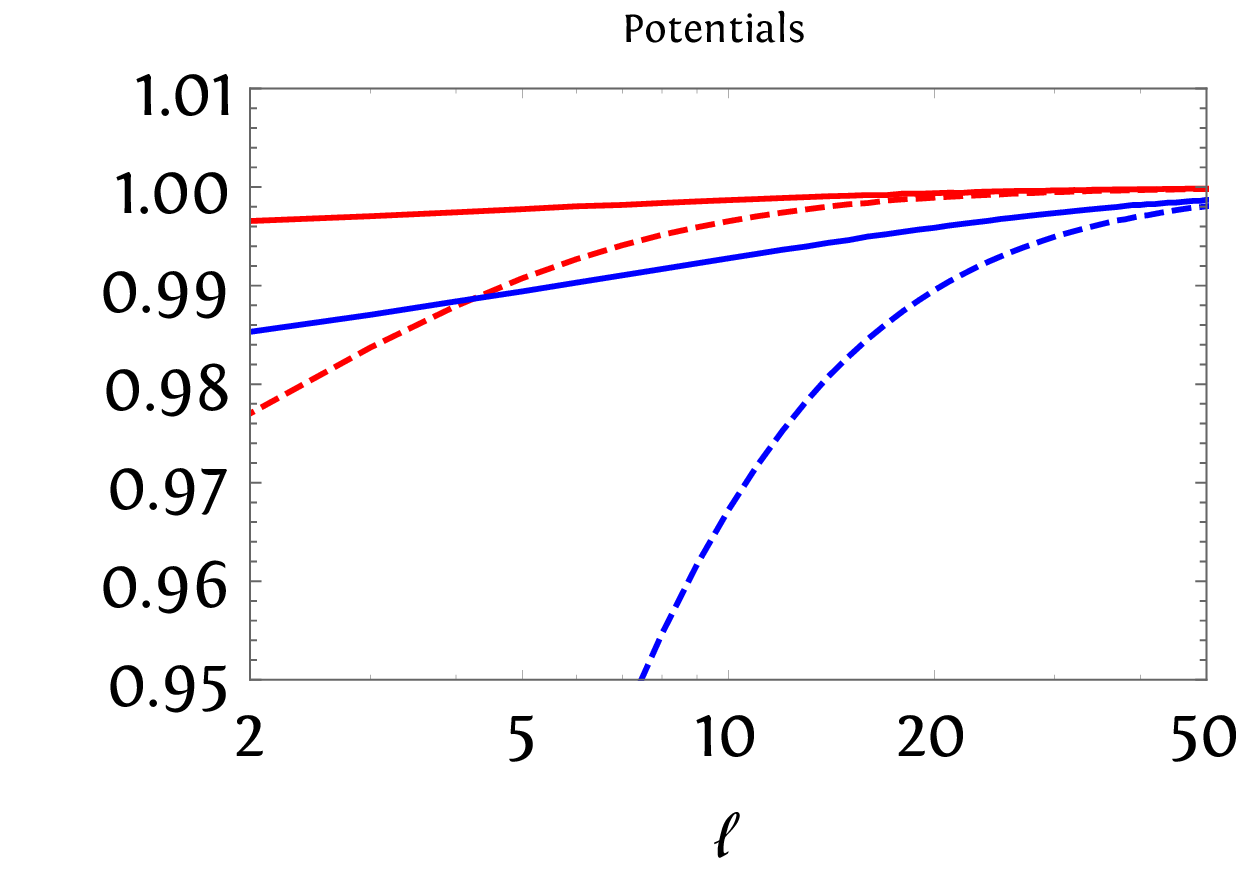}
\caption{Ratios of angular power spectra including some kind of GR corrections to the standard Newtonian approximation, $C_\ell^{\rm Newt}$. Red (blue) curves refer to a shallow (deep) survey, while solid (dashed) lines to narrow (broad) source redshift distributions. (See text for details.)}\label{fig:C_l-GR}
\end{figure}

To give a flavour of how GR corrections alter the Newtonian prediction, we show in Fig.~\ref{fig:C_l-GR} the ratio between a galaxy clustering angular power spectrum in which some GR effect has been switched on and the simple Newtonian result. We factorise the various corrections as follows: RSDs (top, left panel); velocity (top, right panel), where we include terms proportional to the velocity along the line of sight; weak lensing convergence (bottom, left panel); and potentials (bottom, right panel), which account for the effect of gravitational potentials at the source, SW, ISW and time-delay terms \citep[see e.g.][]{Challinor:2011bk}. The two sets of curves refer to different surveys, all of them assumed to observe a certain population of galaxies with constant bias 1.5, distributed according to a Gaussian window centred at redshift $z_m$ and with width $\sigma_w$. Specifically, we compare a shallow survey with $z_m=0.5$ (red curves) to a deep survey with $z_m=2.0$ (blue curves). Moreover, we consider both the cases of a narrow (solid lines) and a broad (dashed lines) window, viz.\ $\sigma_w=0.01$ and $0.1$ for the shallow survey and $\sigma_w=0.05$ and $0.5$ for the deep survey. From Fig.~\ref{fig:C_l-GR}, it is apparent that, although RSDs and lensing are the dominant, the other terms can also become important, particularly for as deep a survey as the SKA.

The GR corrections to the power spectrum should be observable for certain SKA surveys, using the multi-tracer technique---the task of forecasting this capacity is under way. Even if these corrections are not directly detectable, it is essential to include them in theoretical analysis of the power spectrum, for two related reasons:
\begin{enumerate}
\item to avoid biasing parameter estimations through a theoretical systematic of incorrect modelling;
\item to correctly extract maximal information from the largest scales.
\end{enumerate}

\subsection{Primordial Non-Gaussianity}\label{ssec:ng}
The largest cosmic scales are a crucial source of information about the physical processes at play during the early stages of the Universe's evolution. The standard model of inflation and its generalisations predict seed primordial density fluctuations with some level of non-Gaussianity in the probability distribution \citep[e.g.][]{Bartolo:2004if}. We can parameterise the non-Gaussianity  in Bardeen's gauge invariant potential $\Phi$ as the sum of a linear Gaussian term $\phi$ and a quadratic correction \citep{Verde:1999ij,Komatsu:2001rj}, i.e.\
\begin{equation}
\Phi=\phi+\fnl\left(\phi^2-\langle\phi^2\rangle\right). \label{lpng}
\end{equation}
If the distribution of primordial density perturbations is not Gaussian, it cannot be fully described by a power spectrum; we rather need higher-order moments such as the bispectrum. In particular, different models of inflation give rise to different bispectrum shapes, thus making the study of PNG valuable for obtaining a deeper knowledge of the physics of inflation.

The standard single-field inflationary scenario generates negligibly small deviations from Gaussianity. These deviations are said to be of the local shape, and the related bispectrum of Bardeen's potential is maximised for squeezed configurations, where one of the three wavenumbers has a much smaller magnitude than the other two. In this case, the PNG parameter, $\fnl$, is expected to be of the same order as the slow-roll parameters, namely very close to zero \citep{Falk:1992sf}. However, this does not mean that $\fnl \approx 0$. Due to the inherent non-linearity of GR, it is not possible for Gaussianity in the primordial curvature perturbation to be reflected exactly in the density perturbation. A non-linear GR  correction to the initial conditions leads to an effective $f_{\rm NL}^{\rm GR}=-5/3$ in large-scale structure \citep{Verde:2009hy}.

Local-shape PNG can also be generated when an additional light scalar field other than the inflaton contributes to the observed curvature perturbations \citep{Bartolo:2004if}. This happens for instance in curvaton models \citep{Sasaki:2006kq,Assadullahi:2007uw} or in multi-field models \citep{Bartolo:2001cw,Bernardeau:2002jy,Huang:2009vk}. Other than local-type PNG, there are inflationary models in which the kinetic term of the inflaton Lagrangian is  non-standard, containing higher-order derivatives of the field itself. One significant example of this is the `DBI' model \citep{Alishahiha:2004eh,ArkaniHamed:2003uz,Seery:2005wm}, where the primordial bispectrum is maximised for configurations where the three wavevectors have approximately the same amplitude---the so-called equilateral-type PNG \citep{Creminelli:2006rz}. Otherwise, for deviations from Gaussianity evaluated in the regular Bunch-Davies vacuum state, the primordial potential bispectrum is of local or equilateral shape, depending on whether or not 
higher-order derivatives play a significant r\^ole in the evolution of the inflaton field. If the Bunch-Davies vacuum hypothesis is dropped, the resulting bispectrum is maximal for squashed configurations \citep{Chen:2006nt,Holman:2007na}. Lastly, another shape of the bispectrum can be constructed that is nearly orthogonal to both the local and equilateral forms \citep{Senatore:2009gt}.

The best probe of PNG up to now has relied on measuring the CMB temperature anisotropy bispectrum \citep[see][for the latest results]{Ade:2013ydc}. However, it has been demonstrated that PNG also induces an additional, peculiar scale and redshift dependence in a biased tracer of the underlying matter distribution \citep{Dalal:2007cu,Matarrese:2008nc,Schmidt:2010gw,Desjacques:2011jb}. The modification $\Delta b_X(z,k)$ to the Gaussian large-scale bias $b_X$ of a biased tracer $X$ induced by local-type PNG is
\begin{equation}
\Delta b_X(z, k) =b_X(z)+3[b_X(z)-1]\frac{\Omega_{m}H_0^2\delta_c}{k^2T(k)D_+(z)}\fnl,\label{eq:ng-bias}
\end{equation}
where $b_X(z)$ is assumed scale-independent, $\delta_c$ is the critical value of the matter over-density at collapse, the transfer function $T(k)\to 1$ on large scales, and $D_+(z)$ is the linear growth factor of density perturbations normalised to unity today.\footnote{Note that we adopt the large-scale structure (LSS) convention, which implies $f_{\rm NL}^{\rm LSS} \approx 1.3 f_{\rm NL}^{\rm CMB}$ \citep[see e.g.][for more details]{Dalal:2007cu,Afshordi:2008ru}.}

It is clear from this equation that the effect of PNG on the power spectrum grows on large scales, as $k^{-2}$. These are the same scales on which GR corrections are becoming significant. Therefore, one needs to incorporate the GR corrections in theoretical analysis in order to make accurate (and non-biased) predictions and estimates of PNG \citep{Camera:2014sba}. Similarly, an incorrect treatment of PNG on scales where its effects are not negligible may threaten future cosmological experiment accuracy \citep[e.g.][]{Camera:2014dia}.

\subsection{Modified Gravity}\label{ssec:mg}
Einstein's theory of GR has been tested to exquisite precision on `small' scales, namely in the laboratory, the Solar System and with the help of pulsars. On the contrary, tests on scales approaching the cosmological horizon are still rather weak. However, there is great interest in testing GR on very large scales in the context of DE, as the cosmological constant suffers from a plethora of theoretical problems. Since the accelerated expansion of the Universe is a late-time phenomenon, it is natural to look for hints concerning the physical nature of the underlying mechanism on very large scales.

The landscape of MG theories is rather heterogeneous, and there is little reason to prefer any one model over the others. Attempts to parameterise this landscape in a suitably generic way have been only partially successful \citep{PhysRevD.76.104043,Linder:2007hg,Baker:2012zs,Battye:2012eu} although methods based on effective field theory approaches provide a possible avenue for the large class of scalar field models with second-order equations of motion, like Horndeski-type theories \citep{Piazza:2013coa,Bellini:2014fua}. Phenomenological parameterisations can in principle describe the full range of deviations from GR relevant for cosmology and thus fully exploit the information contained in the data by effectively modelling the metric perturbations \citep[see][and references therein]{Kunz:2012aw}. However, phenomenological models are mostly useful to capture the evolution of linear perturbations, which effectively limits them to large scales.

Many MG theories modify the expansion history of the Universe---often by construction, as they are intended as alternatives to DE. However, it has been shown that many DE models are also able to produce arbitrary $w(z)$, and MG models can be tuned to mimic the cosmological constant value of $w(z)=-1$. Therefore, the expansion history is not a smoking gun for MG. On the other hand, MG theories in general also alter the growth history of the Universe, i.e.\ the evolution of the metric perturbations and thus the growth of the matter density contrast too. Many MG models modify the effective strength of gravity as a function of scale and/or redshift, break the equivalence principle and so forth, thus yielding a number of possible observational signatures. On large (linear) scales, the linear growth rate, $f(z)$, is the most sensitive probe of the growth history, and can be well-measured using RSDs from galaxy redshift surveys and IM \citep[see][for IM]{Hall:2012wd}. Another important quantity in this context is 
the scalar anisotropic stress (or gravitational slip), which is generally non-zero in MG theories, especially on very large scales \citep{Saltas:2010tt}. The anisotropic stress is an attractive test for deviations from the standard model because it is measurable without any assumptions on the initial power spectrum or the bias by combining peculiar velocities (e.g.\ RSDs from IM) with weak lensing measurements \citep{Motta:2013cwa}. Its presence appears to be also linked directly to a modification of the propagation of gravitational waves, which can be used as a way to define what MG means \citep{Saltas:2014dha,Camera:2013xfa}.

Going to very large scales is important for many reasons. Generally speaking, the small scales where perturbations are highly non-linear tend to be less useful for cosmological tests of GR, as MG theories generically need to be screened on small scales to avoid violating the strong Solar System constraints (although the mildly non-linear region is interesting). In addition, accurate predictions on non-linear scales require dedicated $N$-body simulations for each theory, and as mentioned above, it is unclear how to use the unifying phenomenological framework on non-linear scales. Baryonic effects on small scales are also badly understood and add a systematic error which effectively renders these scales unusable for precision cosmology for the time being. Then, as the effects that we are looking for are small, we need as many modes as we can get, which also pushes us to large volumes. Last but not least, the scale dependence of deviations from GR is a crucial observable to distinguish between different models, 
and first we need a wide range of scales to be able to observe a scale dependence, and secondly the horizon scale is a natural place where to look for this effect.

Eventually, another way to probe MG at early times provides a link to PNG. According to \citet{Bartolo:2014kaa}, modifications of GR during inflation can create a non-zero quasi-local bispectrum with a non-negligible amplitude that can be probed with the methods described in this Chapter. Furthermore, the GR effects that become relevant on very large scales are actually beneficial for testing gravity on those scales, as they contain additional information about the evolution of perturbations. How to extract this additional information best is currently an active research topic, but the SKA is uniquely suited to  take advantage of this new source of constraints.

\section{Accessing Ultra-Large Cosmic Scales}\label{sec:uls2}
In this Section, we review the two envisaged techniques that the SKA will exploit to tackle the difficult problem of accessing the largest cosmic scales, viz.\ IM (both from and after the EoR) and galaxy multi-tracing. In Sec.~\ref{ssec:systematics}, we also discuss the most important systematics that we shall have to deal with for a fruitful exploitation of such methods. As a figure of merit, we adopt the achievable accuracy $\sigma(\fnl)$ on a forecast measurement of the PNG parameter. However, the techniques presented here will in general enable us to investigate all the peculiar phenomena which occur on extremely large cosmic scales. For example, dedicated analyses aiming at studying other ultra-large scale effects such as GR corrections are being performed at the time of the editing of this Chapter.

\subsection{Intensity Mapping After the Epoch of Reionisation}\label{ssec:im}
IM is an alternative approach for probing the density field of the large-scale structure \citep{Battye:2004re,2008MNRAS.383..606W,Chang:2007xk,Peterson:2009ka}. It involves mapping out the combined emission of the $21$ cm, or HI, line from unresolved galaxies. In doing so, the large-scale structure is detected in three dimensions. If one foregoes identifying individual galaxies, one can greatly speed up the observation and detection of the large-scale structure. IM experiments are sensitive to structures at a redshift range that is observationally difficult to span for ground-based optical surveys \citep{Seo:2009fq}. SKA IM surveys will cover a large fraction of the sky (around 25,000 sq. deg.) over an extremely broad redshift range, making it possible to access the largest cosmic scales in the late Universe. Here, we outline the basics of late-time IM, but refer to \citet{Santos:SKA14} for a more detailed description.

In order to model the 21 cm power spectrum, it is crucial to quantify the bias function between the matter and HI power spectra, $b_{\rm HI}(k,z)$, which allows us to write
\begin{equation}
P^{\rm HI}(k,z)=b^2_{\rm HI}(k,z)\,P^\delta(k,z),
\end{equation}
where $P^\delta(k,z)$ is the underlying matter power spectrum. Several techniques can be adopted in order to extract mock HI power spectra from realistic hydrodynamic or $N$-body simulations of the cosmic large-scale structure. In one of the simplest approaches \citep{bagla10,sarkar,bhara}, the HI is assigned to particles that belong to dark matter haloes as identified in the simulated cosmological volume and the HI content of each halo of mass $M$, $M_{\rm HI}(M)$, needs to be characterised. Other, more recent approaches are instead based on a particle-by-particle method where the HI is assigned to gas particles according to more refined physical prescriptions which take into account self-shielding effects and the conversion to molecular hydrogen \citep[e.g.][]{dave,Popping:2009ym,rahmati,duffy,mao}. A comprehensive comparison of several different methods in terms of 21 cm power spectrum is redshift space, performed in the post-reionisation era at $z\simeq2-4$, has been recently presented by \citet{villa14},
 who used high resolution cosmological hydrodynamic simulations. They show that the $b_{\rm HI}(k,z)$ approaches a constant value to a good approximation at scales $k\lesssim1h$ Mpc$^{-1}$, that galactic feedback in the form of winds is not affecting the signal on the largest scales and that the signal is dominated at least at $z<6$ by HI residing in haloes. Although a full characterisation of the bias function in the $(k,z)$-plane is difficult to make, it is reassuring that on extremely large scales the overall HI bias is flat, thus allowing us to perform cosmological studies of MG and PNG. It is also worth mentioning that the other key quantity is the total HI fraction, $\Omega_{\rm HI}$ \citep[see][]{Padmanabhan:2014zma}, which is in fact measured by observations of quasar absorption lines of the Lyman-$\alpha$ forest and damped Lyman-$\alpha$ systems at $z>2$ \citep{tescari,
Zafar:2013bha}. By means of this data we can then constrain the HI bias.

\subsubsection{Primordial Non-Gaussianity Probed by Scale-Dependent Bias}
\citet{Camera:2013kpa} explored this method for constraining the PNG of primordial fluctuations. IM experiments seem ideally suited for this goal \citep[see][for proposals for doing so in the EoR]{Joudaki:2011sv,Hazra:2012qz,D'Aloisio:2013sda,Lidz:2013tra,Mao:2013yaa}. As argued before, the tightest constraints on $\fnl$ will be obtained for large and deep surveys. Therefore, the volume of the survey determines the ability to probe below $\fnl$ of $\mathcal O(1)$. For such a method to be successful, we need a deep survey with a large bandwidth accessing frequencies of $400$ MHz and below. Assuming that line-of-sight scattering and self-absorption phenomena can be neglected, the HI line radiation can be related to the differential number counts of halo objects \citep[e.g.][]{Challinor:2011bk}, from which we can estimate $b_\mathrm{HI}$. Crucially, given our fundamental ignorance about the redshift evolution of the bias, we need to span a wide range of redshifts to capture $b_\mathrm{HI}(k,z)\neq 1$. This is 
because, as it is clear from Eq.~\eqref{eq:ng-bias}, no PNG effect is to be detected if the tracer we are looking at is unbiased with respect to the underlying matter density distribution.

\begin{figure}
\centering
\includegraphics[width=0.75\textwidth]{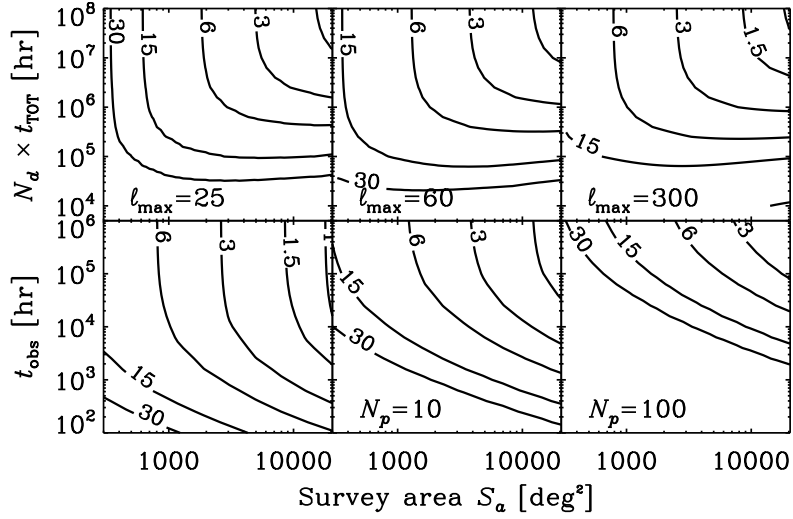}
\caption{Forecast 1$\sigma$ marginal error contours on $f_{\rm NL}^{\rm loc}$ as a function of surveyed sky and total observation time, for a dish survey with $N_d$ dishes (upper panels) and an interferometer making $N_p$ pointings (lower panels) \citep[from][]{Camera:2013kpa}.}\label{fig:exp}
\end{figure}

Fig.~\ref{fig:exp} shows $\sigma(f_{\rm NL}^{\rm loc})$ contours in the plane of the surveyed area and total observation time obtained with an HI IM experiment with bandwidth between 250 and 1000 MHz, corresponding to $0.5\lesssim z\lesssim4.5$, subdivided into $75$ frequency bins of width $\Delta\nu=10$ MHz \citep{Camera:2013kpa}. Abscissas roughly cover from a $15\times15\,\mathrm{deg}^2$ survey to half sky. The three top panels stand for the dish survey case, where the $y$-axis actually shows total observation time, $t_\mathrm{TOT}$, multiplied by the number of dishes, $N_d$. Three maximum angular modes are presented, $\ell_\mathrm{max}=25$, $60$, and $300$, corresponding to dish diameters of $5$, $15$ and $80$ m at $z\approx3$.
For higher angular resolution, interferometers may be a better option. In the bottom panels of Fig.~\ref{fig:exp}, we show $\sigma(f_{\rm NL}^{\rm loc})$ for such a possibility using $1$, $10$ and $100$ pointings. Choosing $D_a\simeq80$ m as the diameter for the array, the resolution is set at $\ell_\mathrm{max}\simeq300$. Here, the main design parameter is the field of view, $\fov$, which sets $\ell_\mathrm{min}=2\pi/\sqrt{\fov}$ and is fixed by the effective size of each element, $d_\mathrm{eff}\sim \lambda/\sqrt{\fov}$. For a `dense array', this is related to the number of elements, $N_e\sim D_a^2/d_\mathrm{eff}^2$. Given that the maximum angular scale is set by the $\fov$,  by adding more pointings, we simply diminish the signal variance by $N_p$, though the noise increases too, because the observation time $t_\mathrm{obs}\to t_\mathrm{obs}/N_p$.

SKA1 would use 15 m dishes, viz.\ $\ell_{\rm max}\simeq60$. Specifically, 254 dishes of SKA1-MID, 10,000 hours and a survey area around 30,000 sq. deg. will correspond to $\sigma(f_{\rm NL}^{\rm loc})\sim2$, as can be seen in Fig.~\ref{fig:exp}. On the one hand, SKA1-MID will only go down to 350 MHz, whereas the minimum frequency assumed in Fig.~\ref{fig:exp} is 250 MHz. On the other hand, \citet{Camera:2013kpa} adopted a more conservative value for the system temperature, i.e. 30 K, whilst SKA1-MID will do better, with a system as cold as 20 K. Otherwise, SKA1-SUR will have a band from 900 to 350 MHz ($0.58 < z < 3.06$) and another from 1400 MHz to 700 MHz ($0 < z < 1.1$). Temperature will be slightly higher, at 50 K, for the former and 30 K for the latter. The effective number of dishes would be redshift dependent, most likely $64\times\nu$ for the first band with $(z,\nu) = (0.6,30)$, $(1.0,20)$, $(2.0,9)$ and $(3.0,5)$, whereas for the second band it would be $94\times\nu$ with $(z,\nu)=(0.0, 23)$, $(0.5,
10)$ and $(1.0, 6)$. For interferometric surveys, we would need to wait for SKA phase 2, with the proposed `aperture array' system working below $1\,\mathrm{GHz}$. It should be possible to achieve $\fov\sim 1000\,\mathrm{deg}^2$, thus reaching the $\sigma(f_{\rm NL}^{\rm loc})\lesssim1$ limit. See \citet{Santos:SKA14} for a detailed description of SKA IM specifics.

\subsubsection{Primordial Non-Gaussianity Probed by the Bispectrum}
Compared to observable galaxies, HI is a weaklier biased tracer of the underlying matter distribution, except for the largest scales where the scale-dependent correction due to PNG dominates. Therefore, on moderately large scales, we may model the HI bispectrum as the tree-level matter bispectrum modified by linear and non-linear bias factors, $b_{{\rm HI},1}$ and $b_{{\rm HI},2}$ . Here, we focus on the reduced HI bispectrum in redshift space, $Q^{\rm HI}_s$, which is much less sensitive to other cosmological parameters \citep{2007PhRvD..76h3004S}. We have
\begin{equation}
Q^{\rm HI}_s(k_1,k_2,k_3)=\frac{a_0^{\rm B}(\beta)}{\left[a_0^{\rm P}(\beta)\right]^2} \left[\frac{1}{b_{{\rm HI},1}}Q^{\rm tree}(k_1,k_2,k_3)+\frac{b_{{\rm HI},2}}{(b_{{\rm HI},1})^2}\right],
\end{equation}
where $Q^{\rm tree}$ is the reduced matter bispectrum predicted by the second-order perturbation theory, $a_0^{\rm P}(\beta)=1+2\beta/3+\beta^2/5$ and $a_0^{\rm B}(\beta)=1+2\beta/3+\beta^2/9$, with $\beta\equiv f(z)/b_{{\rm HI},1}$ the linear Kaiser factor.

The relative importance of PNG in the HI bispectrum increases towards higher 
redshifts. This is very promising for measuring the primordial component from the HI bispectrum with SKA1-MID, which covers the whole redshift range from $z\sim0$ to $z\sim3$ ($350-1420$ MHz). In addition, the large survey volume enabled by IM would result in a much better constraint on $f_{\rm NL}$. The full SKA, with its lower system temperature, better u-v coverage, and broader frequency band (hence broader redshift coverage and larger survey volume) will provide us with more stringent constraints on PNG than those achievable by other experiments such as the Tianlai project.

For SKA1-MID, IM may be conducted with the dishes used individually in auto-correlation mode, being calibrated using interferometry. In this case, $k_{\rm max}$ is limited by the Nyquist frequency, as well as the smallest scale above which we could trust the tree-level matter bispectrum. We adopt a non-linear scale cutoff, $k_{\rm nl}$, for the tree-level matter bispectrum, by requiring the variance in the density contrast field at $\pi/(2k_{\rm nl})$ to equal $0.5$ in each redshift bin. With $N_d=254$ $15$ m diameter dishes, a survey area of 20,000 ${\rm deg}^2$, and a total integration time of 5,000 hr, we find $\sigma(f_{\rm NL}^{\rm loc})=45.7$ and $\sigma(f_{\rm NL}^{\rm eq})=214.3$ when we marginalise over the HI bias factors $b_{{\rm HI},1}$ and $b_{{\rm HI},2}$ at each redshift bin, while $\sigma(f_{\rm NL}^{\rm loc})=15.3$ and $\sigma(f_{\rm NL}^{\rm eq})=61.8$ if we assume a constant bias factors. Otherwise, by using interferometry with the full SKA, $k_{\rm max}$ is set by the non-linear scale $k_
{\rm nl}$, and  we find much accurate marginalised errors $\sigma(f_{\rm NL}^{\rm loc})=6.6$ and $\sigma(f_{\rm NL}^{\rm eq})=55.4$, or  $\sigma(f_{\rm NL}^{\rm loc})=2.2$ and $\sigma(f_{\rm NL}^{\rm eq})=10.9$.

\subsection{Intensity Mapping from the Epoch of Reionisation}
The PNG affects the clustering of the early star-forming galactic haloes responsible for creating a network of ionised patches in the surrounding intergalactic medium during the EoR. This leaves a PNG imprint on the HI tomographic mapping in the intergalactic medium using its redshifted 21 cm radiation. On large scales, where the typical size of ionised regions is much smaller than the scale of interest, we can neglect the non-linear effects of reionisation patchiness on the 21 cm power spectrum. Then the 21 cm temperature power spectrum during the EoR can be written as
\begin{equation}
P^{\Delta T}({\bf k},z) = \widetilde{\delta T}_b^2\bar{x}_{\rm HI}^2\left[ b_{\rm HI}(k,z) + \mu_{\bf k}^2 \right]^2P^\delta(k,z),
\end{equation}
where $\widetilde{\delta T}_b (z) = 23.88\,(\Omega_{\rm b}h^2/0.02) \sqrt{0.15/(\Omega_mh^2)(1+z)/10}$ mK, $\bar{x}_{\rm HI}(z)$ is the global neutral hydrogen fraction, and $\mu_{\bf k}\equiv {\bf k} \cdot\hat{\bf n}/k$, i.e.\ the cosine of angle between the line-of-sight $\hat{\bf n}$ and wave vector ${\bf k}$ of a given Fourier mode.

The ionised density bias $b_{\rm HII}$ is the fundamental quantity derived from reionisation models, related to the neutral density bias $b_{\rm HI}$ by $b_{\rm HI} = \left(1-\bar{x}_{\rm HII}\,b_{\rm HII}\right)/\bar{x}_{\rm HI}$. The reionisation in the presence of PNG can be modelled using two independent methods as follows \citep{D'Aloisio:2013sda}:
\begin{enumerate}
\item \textit{Excursion-set model of reionisation (ESMR)} -- We can use a parameter $\zeta_{\rm ESMR}$ to characterize the efficiency of the local collapsed fraction of mass in luminous sources above some mass threshold in releasing ionising photons into the intergalactic medium \citep{Furlanetto:2004ha}. The full functions $\bar{x}_{\rm HI}(z)$ and $b_{\rm HII}(k,z)$ are set by two parameters, $f_{\rm NL}^{\rm loc}$ and $\zeta_{\rm ESMR}$.
\item \textit{Phenomenological model} -- Similar to the scale-dependent halo bias, \cite{D'Aloisio:2013sda} derived a scale-dependent non-Gaussian correction to the ionised density bias, $\Delta b_{\rm HII}(k,z)$, analogous to Eq.~\eqref{eq:ng-bias}. It depends upon the scale-independent Gaussian ionised density bias $b_{\rm HII}(z)$. Therefore, we can marginalise $\fnl$ over two phenomenological parameters, $\bar{x}_{\rm HI}(z_i)$ and $b_{\rm HII}(z_i)$, in each redshift bin $z_i$.
\end{enumerate}
Both methods can be used to constrain PNG with the 21 cm power spectrum from the EoR. \citet{Mao:2013yaa} demonstrated that for a single frequency bin measurement, their constraints on $f_{\rm NL}^{\rm loc}$ are just as good. However, a model such as the ESMR can be used to combine multiple frequency bin measurements because it fixes the reionisation history (and therefore $P^{\Delta T}$) for a given value of $f_{\rm NL}$ and $\zeta_{\rm ESMR}$. Thus, it appears more promising. 

For a numerical evaluation, we assume that residual foregrounds can be neglected for $k_\parallel \ge k_{\parallel,{\rm min}}=2\pi/(yB)$, e.g.\ $k_{\parallel,{\rm min}}= 0.055\,{\rm Mpc}^{-1}$ at $z=10.1$. The minimum $k_\perp$ is set by the minimum baseline, $k_{\perp,{\rm min}} = 2\pi L_{\rm min}/(\lambda D_a)$. We focus on the 21 cm signal on large scales $k_{\rm max} = 0.15$ Mpc$^{-1}$, so that linearity conditions are met. We assume we can combine information from 7 frequency bins, each of 6 MHz bandwidth, covering the evolution of ionised fraction $\bar{x}_{\rm HII} \approx 0.06 - 0.70$, corresponding to $z \approx 9.5 - 13.4$ in the ESMR method. Using the Fisher matrix formalism, we forecast that the SKA1-LOW can constrain the local-type PNG, $\sigma(f_{\rm NL}^{\rm loc}) = 7.4$, using the ESMR method. Furthermore, if the sensitivity can be improved by 4 times better, then the full SKA can put as tight a constraint as $\sigma(f_{\rm NL}^{\rm loc}) = 4.7$ by using the 21 cm power spectrum from the EoR 
alone.

\subsection{Multi-Tracer Technique}\label{ssec:tracers}
Measuring clustering statistics on large scales has major the advantage of probing non-Gaussian effects on halo/galaxy bias where they are stronger. Unfortunately though, those same scales are limited by cosmic variance, i.e.\ the lack of enough independent measurements for the scales we are trying to probe, given the limited size of the volume that is observed. \citet{Seljak:2008xr} proposed a way to get around cosmic variance limitation by using different biased tracers of the underlying matter distribution. The main idea is that with at least two differently biased tracers, one can measure the ratio of these two biases to a level that is only limited by shot noise, hence beating cosmic variance. This is particularly sensitive when the bias of one tracer is much larger than that of the other. In order to understand this, let us assume that we measure two density maps at a given redshift, one for a biased tracer and another for the dark matter itself. Due to cosmic variance, there will be several 
cosmologies that are consistent with the dark matter map. However, the ratio of the two maps should give a direct measurement of the bias, with an uncertainty just given by the shot noise of the tracer \citep[see also e.g.][]{Abramo:2013awa}.

\begin{figure}
\centering
\includegraphics[width=0.75\textwidth]{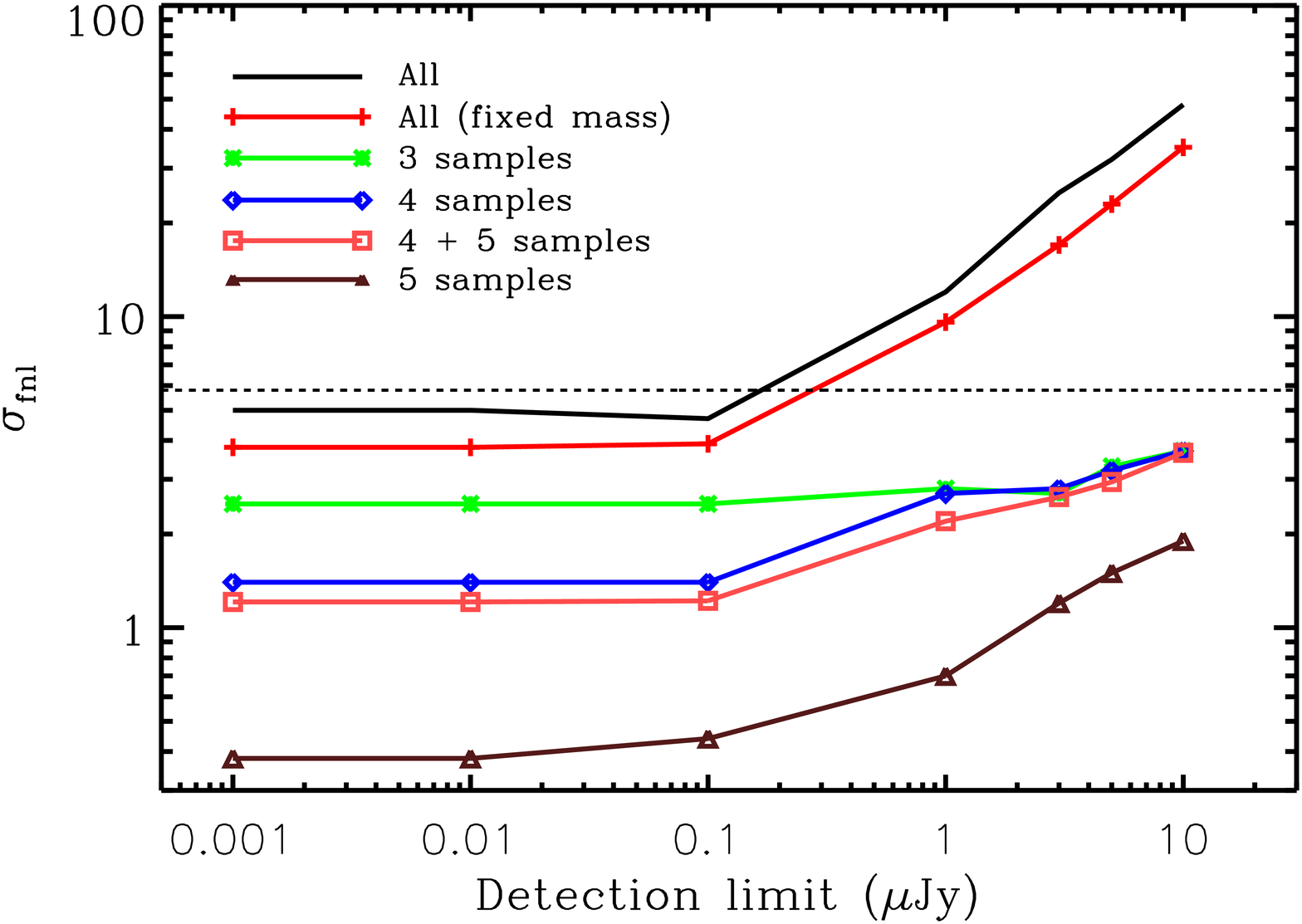}  
\caption{Constraints on $f_{\rm NL}^{\rm loc}$ obtained with the multi-tracer method as a function of the flux cut used to detect galaxies \citep[from][]{Ferramacho:2014pua}. The horizontal line represents the best constrain obtained by the Planck collaboration.}\label{fig:tracers}
\end{figure}

\citet{Ferramacho:2014pua} developed this in the context of radio continuum surveys for SKA and its pathfinders. They explored the strong correlation that exists between radio galaxy types and their bias, which can be at first order linked to a single halo mass for each type. This seems to hold true for most types of radio galaxies, such as radio-quiet and radio-loud active galactic nuclei, including FRI and FRII galaxies, and also star-forming galaxies, including starbursts. Using simulated catalogues, and taking into account all the statistical information from continuum galaxy surveys---namely the combined auto- and cross-correlation angular power spectra of multi-tracer galaxies---\citet{Ferramacho:2014pua} showed that it is possible to strongly reduce the impact of cosmic variance at large scales. By employing the Fisher matrix formalism, they forecast that this method will constrain the local-type PNG parameter up to an amazing accuracy of $\sigma(f_{\rm NL}^{\rm loc})\sim 0.7$, in the most optimal 
case where all the galaxy populations may be distinguished over the whole redshift range, and $\sigma(f_{\rm NL}^{\rm loc})\sim 2.9$, in a much more conservative framework accounting for realistic limitations to the observational identification of each galaxy population.\footnote{For more insight about the impact of redshift information for radio continuum surveys in the context of the SKA, see also \citet{Camera:2012ez} and \citet{Raccanelli:2014kga}, where in the latter work they used the ISW effect and redshift measurements from other experiments to constrain $\sigma(f_{\rm NL}^{\rm loc})\sim1$. Comparable, encouraging results for SKA HI galaxy redshift surveys have recently been obtained by \citet{Camera:2014bwa}.} Note that the improvement obtained through the multi-tracer analysis is indeed significant, as the whole galaxy catalogue without any galaxy type differentiation only allows for constraints on $\fnl$ with an error of $\sigma(f_{\rm NL}^{\rm loc})=32$ in the realistic scenario. Fig. \ref{fig:tracers} resumes the forecast constraints as a function of flux cut limit \citep[see][for details]{Ferramacho:2014pua}.

\subsection{Systematics Occurring on the Largest Scales}\label{ssec:systematics}
The SKA should be able to provide a statistical detection of large scale effects due to its high sensitivity and large volumes probed, it will thus be crucial to control the systematics which will occur on these same scales. For the measurements we have been discussing, the following contaminants will need to be considered:
\begin{itemize}
\item \textit{Masks} -- Several systematic effects related to partial sky coverage can result in biases and spurious large-scale signals. This is an issue for both galaxy surveys as well as IM. This effect can in principle be dealt with by means of inversion methods, as has been done by the CMB community for instance to address the masking of our own galaxy.
\item \textit{Foreground subtraction} -- For IM experiments, most foreground removal algorithms subtract out modes that are smoothly-varying in frequency, since this is how most galactic and extra-galactic foregrounds behave. In the process, some fraction of large scale cosmological modes along the line of sight will also be subtracted. The effect should be most important for scales around the total bandwidth, but there will be a contamination on smaller scales too \citep{Alonso:2014sna}.
\item \textit{Correlated noise} -- Auto-correlation observations are needed to access the largest angular scales with IM using the SKA. There is a noise term that has correlations in time, e.g.\ a non-white noise behaviour, with a drift that depends on the receiver design. This term cancels out in interferometric observations but is present in the auto-correlation signal from each dish. Therefore, strategies will have to be designed to remove this contribution. The situation can be particularly critical for large scale modes since this contamination shows up on time scales larger than the typical "1/f" knee of the receiver, which can translate into striping artefacts on large angular scales. The usual solution is to design fast survey strategies capable of probing the required angles over a period that is shorter than the typical time-scale of the receiver. This will impose strong constraints on the receiver design that need to be accessed. Nevertheless other techniques should be explored. For instance the 
fact that the noise can also be a smooth function of frequency may allow for a solution in dealing with this effect.
\item \textit{Mosaicking} -- Large scale surveys will require several pointings (observations in different directions) in order to cover the required sky area---the so called `mosaicking'. This is true for any of the surveys, either continuum, HI galaxy or HI intensity mapping (the situation with SKA1-LOW is different due to its large field of view). Careful design of the scanning strategy needs to be made so that the noise is sufficiently uniform across the sky, in particular when overlapping the pointings. For example, fluctuations of this noise can generate different flux sensitivities across the sky and affect the number of detected galaxies in a threshold survey. This could in turn generate spurious fluctuations on angular scales (the situation is less of a concern if we consider a high flux cut in the galaxy survey, such as 10$\sigma$). The same issue can arise if there are non-negligible fluctuations on the calibrated signal. These problems can in principle be  dealt with through the mosaicking 
approach as well as an accurate calibration process. Moreover, these fluctuations should be uncorrelated on large angular scales which should allow them to be distinguished from the signal.
\end{itemize}

\section{Conclusions}\label{sec:conclusions}
The access to observations on ultra-large scales that are possible with the SKA enables a range of unique science goals at the forefront of cosmology. They include constraints on DE and MG, amongst the key science goals identified by the \citet{national2010New}. The constraints possible with the SKA will rival those the Euclid satellite will deliver \citep{EditorialTeam:2011mu,Amendola:2012ys}, but will be complementary especially where the systematic effects are concerned. The SKA will also be able to use relativistic effects to test GR on the largest scales, and it will surpass the constraints on PNG obtained by the Planck satellite. The SKA will probe the inflationary dynamics that took place in the Universe just fractions of a second after the Big Bang, and it will also test whether gravity was modified during that epoch.

This Chapter, together with others in this review, clearly shows that the SKA will be able to probe the foundations of our cosmological models, the evolution of the Universe and the nature of its constituents. Moreover, as we have described, this will be accessed by a range of SKA surveys which will probe the underlying signal in different ways. The SKA will contribute in unique ways to broaden our understanding of the cosmos, and the SKA data will become a key tool for cosmology in the 21st century.

\section*{Acknowledgements}SC acknowledges support from the European Research Council under the EC FP7 grant No. 280127 and FCT-Portugal under Post-Doctoral Grant No.\ SFRH/BPD/80274/2011. AR is supported by the Templeton Foundation. Part of the research described in this paper was carried out at the Jet Propulsion Laboratory, California Institute of Technology, under a contract with the National Aeronautics and Space Administration. PB is supported by European Research Council grant No.\ StG2010-257080. MS and RM  are supported by the South African Square Kilometre Array Project and the South African National Research Foundation. RM acknowledges support from the UK Science \& Technology Facilities Council (grant ST/K0090X/1). YM was supported within the Labex ILP (reference ANR-10-LABX-63) part of the Idex SUPER, and received financial state aid managed by the Agence Nationale de la Recherche, as part of the programme Investissements d'avenir under the reference ANR-11-IDEX-0004-02. PRS was supported in part by U.S. NSF grant AST-1009799, NASA grant NNX11AE09G, NASA/JPL grant RSA Nos. 1492788 and 1515294, and supercomputer resources from NSF XSEDE grant TG-AST090005 and the Texas Advanced Computing Center (TACC) at The University of Texas at Austin.

\bibliographystyle{apj}
\bibliography{../../Bibliography.bib}

\begin{thebibliography}{86}
\expandafter\ifx\csname natexlab\endcsname\relax\def\natexlab#1{#1}\fi

\bibitem[{Abdalla {et~al.}(2014)}]{Abdalla:SKA14}
Abdalla, F. {et~al.} 2014, Advancing Astrophysics with the Square Kilometre
  Array

\bibitem[{Abramo \& Leonard(2013)}]{Abramo:2013awa}
Abramo, L.~R. \& Leonard, K.~E. 2013, Mon. Not. Roy. Astron. Soc., 432, 318

\bibitem[{Ade {et~al.}(2014)}]{Ade:2013ydc}
Ade, P. {et~al.} 2014, Astron. Astrophys., 571, A24

\bibitem[{Afshordi \& Tolley(2008)}]{Afshordi:2008ru}
Afshordi, N. \& Tolley, A.~J. 2008, Phys. Rev., D78, 123507

\bibitem[{Alishahiha {et~al.}(2004)Alishahiha, Silverstein, \&
  Tong}]{Alishahiha:2004eh}
Alishahiha, M., Silverstein, E., \& Tong, D. 2004, Phys. Rev., D70, 123505

\bibitem[{Alonso {et~al.}(2014)Alonso, Ferreira, \& Santos}]{Alonso:2014sna}
Alonso, D., Ferreira, P.~G., \& Santos, M.~G. 2014, Mon. Not. Roy. Astron.
  Soc., 444, 3183

\bibitem[{Amendola {et~al.}(2013)}]{Amendola:2012ys}
Amendola, L. {et~al.} 2013, Living Rev. Rel., 16, 6

\bibitem[{Arkani-Hamed {et~al.}(2004)Arkani-Hamed, Creminelli, Mukohyama, \&
  Zaldarriaga}]{ArkaniHamed:2003uz}
Arkani-Hamed, N., Creminelli, P., Mukohyama, S., \& Zaldarriaga, M. 2004, JCAP,
  0404, 001

\bibitem[{Assadullahi {et~al.}(2007)Assadullahi, Valiviita, \&
  Wands}]{Assadullahi:2007uw}
Assadullahi, H., Valiviita, J., \& Wands, D. 2007, Phys. Rev., D76, 103003

\bibitem[{{Bagla} {et~al.}(2010){Bagla}, {Khandai}, \& {Datta}}]{bagla10}
{Bagla}, J.~S., {Khandai}, N., \& {Datta}, K.~K. 2010, Mon. Not. Roy. Astron.
  Soc., 407, 567

\bibitem[{Baker {et~al.}(2013)Baker, Ferreira, \& Skordis}]{Baker:2012zs}
Baker, T., Ferreira, P.~G., \& Skordis, C. 2013, Phys. Rev., D87, 024015

\bibitem[{Bartolo {et~al.}(2014)Bartolo, Cannone, Jimenez, Matarrese, \&
  Verde}]{Bartolo:2014kaa}
Bartolo, N., Cannone, D., Jimenez, R., Matarrese, S., \& Verde, L. 2014, Phys.
  Rev. Lett., 113, 161303

\bibitem[{Bartolo {et~al.}(2004)Bartolo, Komatsu, Matarrese, \&
  Riotto}]{Bartolo:2004if}
Bartolo, N., Komatsu, E., Matarrese, S., \& Riotto, A. 2004, Phys. Rept., 402,
  103

\bibitem[{Bartolo {et~al.}(2002)Bartolo, Matarrese, \& Riotto}]{Bartolo:2001cw}
Bartolo, N., Matarrese, S., \& Riotto, A. 2002, Phys. Rev., D65, 103505

\bibitem[{Battye {et~al.}(2004)Battye, Davies, \& Weller}]{Battye:2004re}
Battye, R.~A., Davies, R.~D., \& Weller, J. 2004, Mon. Not. Roy. Astron. Soc.,
  355, 1339

\bibitem[{Battye \& Pearson(2012)}]{Battye:2012eu}
Battye, R.~A. \& Pearson, J.~A. 2012, JCAP, 1207, 019

\bibitem[{Bellini \& Sawicki(2014)}]{Bellini:2014fua}
Bellini, E. \& Sawicki, I. 2014, JCAP, 1407, 050

\bibitem[{Bernardeau \& Uzan(2002)}]{Bernardeau:2002jy}
Bernardeau, F. \& Uzan, J.-P. 2002, Phys. Rev., D66, 103506

\bibitem[{Bertacca {et~al.}(2012)Bertacca, Maartens, Raccanelli, \&
  Clarkson}]{Bertacca:2012tp}
Bertacca, D., Maartens, R., Raccanelli, A., \& Clarkson, C. 2012, JCAP, 1210,
  025

\bibitem[{{Bharadwaj} {et~al.}(2001){Bharadwaj}, {Nath}, \& {Sethi}}]{bhara}
{Bharadwaj}, S., {Nath}, B.~B., \& {Sethi}, S.~K. 2001, Journal of Astrophysics
  and Astronomy, 22, 21

\bibitem[{Bonvin \& Durrer(2011)}]{Bonvin:2011bg}
Bonvin, C. \& Durrer, R. 2011, Phys. Rev., D84, 063505

\bibitem[{Bonvin {et~al.}(2014)Bonvin, Hui, \& Gaztanaga}]{Bonvin:2013ogt}
Bonvin, C., Hui, L., \& Gaztanaga, E. 2014, Phys. Rev., D89, 083535

\bibitem[{Bruni {et~al.}(2012)Bruni, Crittenden, Koyama, Maartens, Pitrou,
  {et~al.}}]{Bruni:2011ta}
Bruni, M., Crittenden, R., Koyama, K., Maartens, R., Pitrou, C., {et~al.} 2012,
  Phys. Rev., D85, 041301

\bibitem[{Camera {et~al.}(2014{\natexlab{a}})Camera, Carbone, Fedeli, \&
  Moscardini}]{Camera:2014dia}
Camera, S., Carbone, C., Fedeli, C., \& Moscardini, L. 2014{\natexlab{a}}

\bibitem[{Camera \& Nishizawa(2013)}]{Camera:2013xfa}
Camera, S. \& Nishizawa, A. 2013, Phys. Rev. Lett., 110, 151103

\bibitem[{Camera {et~al.}(2012)Camera, Santos, Bacon, Jarvis, McAlpine,
  {et~al.}}]{Camera:2012ez}
Camera, S., Santos, M.~G., Bacon, D.~J., Jarvis, M.~J., McAlpine, K., {et~al.}
  2012, Mon. Not. Roy. Astron. Soc., 427, 2079

\bibitem[{Camera {et~al.}(2013)Camera, Santos, Ferreira, \&
  Ferramacho}]{Camera:2013kpa}
Camera, S., Santos, M.~G., Ferreira, P.~G., \& Ferramacho, L. 2013, Phys. Rev.
  Lett., 111, 171302

\bibitem[{Camera {et~al.}(2014{\natexlab{b}})Camera, Santos, \&
  Maartens}]{Camera:2014sba}
Camera, S., Santos, M.~G., \& Maartens, R. 2014{\natexlab{b}}

\bibitem[{Camera {et~al.}(2014{\natexlab{c}})Camera, Santos, \&
  Maartens}]{Camera:2014bwa}
---. 2014{\natexlab{c}}, eprint arXiv:1409.8286

\bibitem[{Challinor \& Lewis(2011)}]{Challinor:2011bk}
Challinor, A. \& Lewis, A. 2011, Phys. Rev., D84, 043516

\bibitem[{Chang {et~al.}(2008)Chang, Pen, Peterson, \& McDonald}]{Chang:2007xk}
Chang, T.-C., Pen, U.-L., Peterson, J.~B., \& McDonald, P. 2008, Phys. Rev.
  Lett., 100, 091303

\bibitem[{Chen {et~al.}(2007)Chen, Huang, Kachru, \& Shiu}]{Chen:2006nt}
Chen, X., Huang, M.-x., Kachru, S., \& Shiu, G. 2007, JCAP, 0701, 002

\bibitem[{{Committee for a Decadal Survey of Astronomy and
  Astrophysics}(National Research Council, 2010)}]{national2010New}
{Committee for a Decadal Survey of Astronomy and Astrophysics}. National
  Research Council, 2010, New Worlds, New Horizons in Astronomy and
  Astrophysics (The National Academies Press)

\bibitem[{Creminelli {et~al.}(2007)Creminelli, Senatore, Zaldarriaga, \&
  Tegmark}]{Creminelli:2006rz}
Creminelli, P., Senatore, L., Zaldarriaga, M., \& Tegmark, M. 2007, JCAP, 0703,
  005

\bibitem[{Dalal {et~al.}(2008)Dalal, Dore, Huterer, \& Shirokov}]{Dalal:2007cu}
Dalal, N., Dore, O., Huterer, D., \& Shirokov, A. 2008, Phys. Rev., D77, 123514

\bibitem[{D'Aloisio {et~al.}(2013)D'Aloisio, Zhang, Shapiro, \&
  Mao}]{D'Aloisio:2013sda}
D'Aloisio, A., Zhang, J., Shapiro, P.~R., \& Mao, Y. 2013, Mon. Not. Roy.
  Astron. Soc., 433, 2900

\bibitem[{{Dav{\'e}} {et~al.}(2013){Dav{\'e}}, {Katz}, {Oppenheimer},
  {Kollmeier}, \& {Weinberg}}]{dave}
{Dav{\'e}}, R., {Katz}, N., {Oppenheimer}, B.~D., {Kollmeier}, J.~A., \&
  {Weinberg}, D.~H. 2013, Mon. Not. Roy. Astron. Soc., 434, 2645

\bibitem[{Desjacques {et~al.}(2011)Desjacques, Jeong, \&
  Schmidt}]{Desjacques:2011jb}
Desjacques, V., Jeong, D., \& Schmidt, F. 2011, Phys. Rev., D84, 061301

\bibitem[{Di~Dio {et~al.}(2013)Di~Dio, Montanari, Lesgourgues, \&
  Durrer}]{DiDio:2013bqa}
Di~Dio, E., Montanari, F., Lesgourgues, J., \& Durrer, R. 2013, JCAP, 1311, 044

\bibitem[{{Duffy} {et~al.}(2012){Duffy}, {Kay}, {Battye}, {Booth}, {Dalla
  Vecchia}, \& {Schaye}}]{duffy}
{Duffy}, A.~R., {Kay}, S.~T., {Battye}, R.~A., {Booth}, C.~M., {Dalla Vecchia},
  C., \& {Schaye}, J. 2012, Mon. Not. Roy. Astron. Soc., 420, 2799

\bibitem[{Falk {et~al.}(1993)Falk, Rangarajan, \& Srednicki}]{Falk:1992sf}
Falk, T., Rangarajan, R., \& Srednicki, M. 1993, Astrophys. J., 403, L1

\bibitem[{Ferramacho {et~al.}(2014)Ferramacho, Santos, Jarvis, \&
  Camera}]{Ferramacho:2014pua}
Ferramacho, L.~D., Santos, M.~G., Jarvis, M.~J., \& Camera, S. 2014, Mon. Not.
  Roy. Astron. Soc., 442, 2511

\bibitem[{Furlanetto {et~al.}(2004)Furlanetto, Zaldarriaga, \&
  Hernquist}]{Furlanetto:2004ha}
Furlanetto, S., Zaldarriaga, M., \& Hernquist, L. 2004, Astrophys. J., 613, 16

\bibitem[{{Guha Sarkar} {et~al.}(2012){Guha Sarkar}, {Mitra}, {Majumdar}, \&
  {Choudhury}}]{sarkar}
{Guha Sarkar}, T., {Mitra}, S., {Majumdar}, S., \& {Choudhury}, T.~R. 2012,
  Mon. Not. Roy. Astron. Soc., 421, 3570

\bibitem[{Hall {et~al.}(2013)Hall, Bonvin, \& Challinor}]{Hall:2012wd}
Hall, A., Bonvin, C., \& Challinor, A. 2013, Phys. Rev., D87, 064026

\bibitem[{Hazra \& Sarkar(2012)}]{Hazra:2012qz}
Hazra, D.~K. \& Sarkar, T.~G. 2012, Phys. Rev. Lett., 109, 121301

\bibitem[{Holman \& Tolley(2008)}]{Holman:2007na}
Holman, R. \& Tolley, A.~J. 2008, JCAP, 0805, 001

\bibitem[{Hu \& Sawicki(2007)}]{PhysRevD.76.104043}
Hu, W. \& Sawicki, I. 2007, Phys. Rev. D, 76, 104043

\bibitem[{Huang(2009)}]{Huang:2009vk}
Huang, Q.-G. 2009, JCAP, 0906, 035

\bibitem[{Jarvis {et~al.}(2014)}]{Jarvis:SKA14}
Jarvis, M.~J. {et~al.} 2014, Advancing Astrophysics with the Square Kilometre
  Array

\bibitem[{Jeong {et~al.}(2012)Jeong, Schmidt, \& Hirata}]{Jeong:2011as}
Jeong, D., Schmidt, F., \& Hirata, C.~M. 2012, Phys. Rev., D85, 023504

\bibitem[{Joudaki {et~al.}(2011)Joudaki, Dore, Ferramacho, Kaplinghat, \&
  Santos}]{Joudaki:2011sv}
Joudaki, S., Dore, O., Ferramacho, L., Kaplinghat, M., \& Santos, M.~G. 2011,
  Phys. Rev. Lett., 107, 131304

\bibitem[{Komatsu \& Spergel(2001)}]{Komatsu:2001rj}
Komatsu, E. \& Spergel, D.~N. 2001, Phys. Rev., D63, 063002

\bibitem[{Kunz(2012)}]{Kunz:2012aw}
Kunz, M. 2012, Comptes Rendus Physique, 13, 539

\bibitem[{Laureijs {et~al.}(2011)Laureijs, Amiaux, Arduini, Augueres,
  {et~al.}}]{EditorialTeam:2011mu}
Laureijs, R., Amiaux, J., Arduini, S., Augueres, J.-L., {et~al.} 2011, ESA-SRE,
  12

\bibitem[{Lidz {et~al.}(2013)Lidz, Baxter, Adshead, \& Dodelson}]{Lidz:2013tra}
Lidz, A., Baxter, E.~J., Adshead, P., \& Dodelson, S. 2013, Phys. Rev., D88,
  023534

\bibitem[{Linder \& Cahn(2007)}]{Linder:2007hg}
Linder, E.~V. \& Cahn, R.~N. 2007, Astropart.Phys., 28, 481

\bibitem[{Mao {et~al.}(2013)Mao, D'Aloisio, Zhang, \& Shapiro}]{Mao:2013yaa}
Mao, Y., D'Aloisio, A., Zhang, J., \& Shapiro, P.~R. 2013, Phys. Rev., D88,
  081303

\bibitem[{{Mao} {et~al.}(2012){Mao}, {Shapiro}, {Mellema}, {Iliev}, {Koda}, \&
  {Ahn}}]{mao}
{Mao}, Y., {Shapiro}, P.~R., {Mellema}, G., {Iliev}, I.~T., {Koda}, J., \&
  {Ahn}, K. 2012, Mon. Not. Roy. Astron. Soc., 422, 926

\bibitem[{Matarrese \& Verde(2008)}]{Matarrese:2008nc}
Matarrese, S. \& Verde, L. 2008, Astrophys. J., 677, L77

\bibitem[{Motta {et~al.}(2013)Motta, Sawicki, Saltas, Amendola, \&
  Kunz}]{Motta:2013cwa}
Motta, M., Sawicki, I., Saltas, I.~D., Amendola, L., \& Kunz, M. 2013, Phys.
  Rev., D88, 124035

\bibitem[{Padmanabhan {et~al.}(2014)Padmanabhan, Choudhury, \&
  Refregier}]{Padmanabhan:2014zma}
Padmanabhan, H., Choudhury, T.~R., \& Refregier, A. 2014, eprint
  arXiv:1407.6366

\bibitem[{Peterson {et~al.}(2009)Peterson, Aleksan, Ansari, Bandura, Bond,
  {et~al.}}]{Peterson:2009ka}
Peterson, J.~B., Aleksan, R., Ansari, R., Bandura, K., Bond, D., {et~al.} 2009,
  astro2010: The Astronomy and Astrophysics Decadal Survey

\bibitem[{Piazza \& Vernizzi(2013)}]{Piazza:2013coa}
Piazza, F. \& Vernizzi, F. 2013, Class.Quant.Grav., 30, 214007

\bibitem[{Popping {et~al.}(2009)Popping, Dave, Braun, \&
  Oppenheimer}]{Popping:2009ym}
Popping, A., Dave, R., Braun, R., \& Oppenheimer, B.~D. 2009, Astron.
  Astrophys., 504, 15

\bibitem[{Raccanelli {et~al.}(2014{\natexlab{a}})Raccanelli, Bertacca, Dore, \&
  Maartens}]{Raccanelli:2013dza}
Raccanelli, A., Bertacca, D., Dore, O., \& Maartens, R. 2014{\natexlab{a}},
  JCAP, 1408, 022

\bibitem[{Raccanelli {et~al.}(2013)Raccanelli, Bertacca, Maartens, Clarkson, \&
  Dore}]{Raccanelli:2013gja}
Raccanelli, A., Bertacca, D., Maartens, R., Clarkson, C., \& Dore, O. 2013,
  eprint arXiv:1311.6813

\bibitem[{Raccanelli {et~al.}(2014{\natexlab{b}})Raccanelli, Dore, Bacon,
  Maartens, Santos, {et~al.}}]{Raccanelli:2014kga}
Raccanelli, A., Dore, O., Bacon, D.~J., Maartens, R., Santos, M.~G., {et~al.}
  2014{\natexlab{b}}, eprint arXiv:1406.0010

\bibitem[{{Rahmati} {et~al.}(2013){Rahmati}, {Pawlik}, {Raicevic}, \&
  {Schaye}}]{rahmati}
{Rahmati}, A., {Pawlik}, A.~H., {Raicevic}, M., \& {Schaye}, J. 2013, Mon. Not.
  Roy. Astron. Soc., 430, 2427

\bibitem[{Saltas \& Kunz(2011)}]{Saltas:2010tt}
Saltas, I.~D. \& Kunz, M. 2011, Phys. Rev., D83, 064042

\bibitem[{Saltas {et~al.}(2014)Saltas, Sawicki, Amendola, \&
  Kunz}]{Saltas:2014dha}
Saltas, I.~D., Sawicki, I., Amendola, L., \& Kunz, M. 2014, Phys. Rev. Lett.,
  113, 191101

\bibitem[{Santos {et~al.}(2014)}]{Santos:SKA14}
Santos, M.~G. {et~al.} 2014, Advancing Astrophysics with the Square Kilometre
  Array

\bibitem[{Sasaki {et~al.}(2006)Sasaki, Valiviita, \& Wands}]{Sasaki:2006kq}
Sasaki, M., Valiviita, J., \& Wands, D. 2006, Phys. Rev., D74, 103003

\bibitem[{Schmidt \& Kamionkowski(2010)}]{Schmidt:2010gw}
Schmidt, F. \& Kamionkowski, M. 2010, Phys. Rev., D82, 103002

\bibitem[{Seery \& Lidsey(2005)}]{Seery:2005wm}
Seery, D. \& Lidsey, J.~E. 2005, JCAP, 0506, 003

\bibitem[{{Sefusatti} \& {Komatsu}(2007)}]{2007PhRvD..76h3004S}
{Sefusatti}, E. \& {Komatsu}, E. 2007, Phys. Rev., 76, 083004

\bibitem[{Seljak(2009)}]{Seljak:2008xr}
Seljak, U. 2009, Phys. Rev. Lett., 102, 021302

\bibitem[{Senatore {et~al.}(2010)Senatore, Smith, \&
  Zaldarriaga}]{Senatore:2009gt}
Senatore, L., Smith, K.~M., \& Zaldarriaga, M. 2010, JCAP, 1001, 028

\bibitem[{Seo {et~al.}(2010)Seo, Dodelson, Marriner, Mcginnis, Stebbins,
  {et~al.}}]{Seo:2009fq}
Seo, H.-J., Dodelson, S., Marriner, J., Mcginnis, D., Stebbins, A., {et~al.}
  2010, Astrophys.J., 721, 164

\bibitem[{{Tescari} {et~al.}(2009){Tescari}, {Viel}, {Tornatore}, \&
  {Borgani}}]{tescari}
{Tescari}, E., {Viel}, M., {Tornatore}, L., \& {Borgani}, S. 2009, Mon. Not.
  Roy. Astron. Soc., 397, 411

\bibitem[{Verde \& Matarrese(2009)}]{Verde:2009hy}
Verde, L. \& Matarrese, S. 2009, Astrophys. J., 706, L91

\bibitem[{Verde {et~al.}(2000)Verde, Wang, Heavens, \&
  Kamionkowski}]{Verde:1999ij}
Verde, L., Wang, L.-M., Heavens, A., \& Kamionkowski, M. 2000, Mon. Not. Roy.
  Astron. Soc., 313, L141

\bibitem[{{Villaescusa-Navarro} {et~al.}(2014){Villaescusa-Navarro}, {Viel},
  {Datta}, \& {Choudhury}}]{villa14}
{Villaescusa-Navarro}, F., {Viel}, M., {Datta}, K.~K., \& {Choudhury}, T.~R.
  2014, eprint arXiv:1405.6713

\bibitem[{{Wyithe} \& {Loeb}(2008)}]{2008MNRAS.383..606W}
{Wyithe}, J.~S.~B. \& {Loeb}, A. 2008, Mon. Not. Roy. Astron. Soc., 383, 606

\bibitem[{Yoo(2010)}]{Yoo:2010ni}
Yoo, J. 2010, Phys. Rev., D82, 083508

\bibitem[{Zafar {et~al.}(2013)Zafar, Peroux, Popping, Milliard, Deharveng,
  {et~al.}}]{Zafar:2013bha}
Zafar, T., Peroux, C., Popping, A., Milliard, B., Deharveng, J.-M., {et~al.}
  2013, Astron. Astrophys.,, A141

\end{thebibliography}

\end{document}